\documentclass[11pt,a4paper]{article}
\usepackage{jheppub}
\usepackage{epsfig}
\usepackage{amsfonts}
\usepackage{amscd}
\usepackage{latexsym}
\usepackage{amsmath,amssymb}
\usepackage{verbatim}
\usepackage{setspace}  
\usepackage{hyperref}
\usepackage{slashed}
\usepackage{float}
\usepackage{marvosym}

\numberwithin{equation}{section}
\def\p{\partial}
\def\kap{\kappa}
\def\ca{{\mathcal A}}

\def\co{{\mathcal O}}

\def\<{\langle }
\def\>{\rangle}
\def\eps{\varepsilon}
\def\gam{\gamma}
\def\half{\frac{1}{2}}

\newcommand{\ba}{\begin{aligned}}
\newcommand{\ea}{\end{aligned}}
\def\beq{\be\begin{array}{c}}
\def\eeq{\end{array}\ee} 
\def\be{\begin{equation}}
\def\ee{\end{equation}}
\def\bea{\begin{eqnarray}}
\def\eea{\end{eqnarray}}

\title{Accelerating Black Hole Thermodynamics with Boost Time}
 
\author[*]{Adam Ball,}
\author[*]{Noah Miller}

\affiliation[*]{Center for the Fundamental Laws of Nature, Harvard University,\\
17 Oxford Street, Cambridge, MA 02138, USA}

\emailAdd{aaball@g.harvard.edu}
\emailAdd{noahmiller@g.harvard.edu}

\abstract{We derive a thermodynamic first law for the electrically charged C-metric with vanishing cosmological constant. This spacetime describes a pair of identical accelerating black holes each pulled by a cosmic string. Treating the ``boost time'' of this spacetime as the canonical time, we find a thermodynamic first law in which every term has an unambiguous physical meaning. We then show how this first law can be derived using Noetherian methods in the covariant phase space formalism. We argue that the area of the acceleration horizon contributes to the entropy and that the appropriate notion of energy of this spacetime is a ``boost mass'', which vanishes identically. The recovery of the Reissner-Nordstrom first law in the limit of small string tension is also demonstrated. Finally, we compute the action of the Euclidean section of the C-metric and show it agrees with the thermodynamic grand potential, providing an independent confirmation of the validity of our first law. We also briefly speculate on the significance of firewalls in this spacetime.}

\begin{document} 
\maketitle
\flushbottom
 
\section{Introduction}

The C-metric describes a pair of accelerating oppositely charged black holes pulled to infinity by cosmic strings \cite{kinwalk}. The event horizons of these black holes are connected by a non-traversable wormhole through which the cosmic string threads \cite{hawkross}. The spacetime has a ``boost" symmetry, much like the boost symmetry of Minkowski space. The C-metric and its close cousin, the Ernst metric, have been used in multiple contexts to explore quantum gravity in spacetimes with a vanishing cosmological constant \cite{swwp, ggsErnst, Dowker:1993bt, Dowker:1994up, hhrErnst, hawkross, maldacena2013cool}. Recently it has been proposed that they hold information about non-perturbative aspects of a putative celestial CFT living on the null infinity of asymptotically locally flat spacetime \cite{stromzhib}.\footnote{The string makes the spacetime not asymptotically flat, but only asymptotically locally flat \cite{AshtekarDray}.}

For any of these applications, it would be beneficial to develop an understanding of the thermodynamics of the C-metric in analogy with the usual laws of black hole thermodynamics. This task is complicated by certain peculiarities of the C-metric. These peculiarities include the lack of a global timelike Killing vector, the existence of a non-compact acceleration horizon, and the presence of the cosmic string.

Our starting point in studying C-metric thermodynamics is to use ``boost time" as the canonical time. That is, our first law will be from the perspective of a static observer.\footnote{The observer cannot see past the horizons, so their perceived universe is only a patch of the full C-metric spacetime, and in particular contains only one of the black holes.} While the boost Killing vector is not globally timelike, adopting this position has proven to be fruitful. After a review of the C-metric and its corresponding background spacetime we discuss how the first law can be derived using the covariant phase space formalism, along the way addressing the subtle issue of the difference between ``global'' and ``local'' first laws. We then show how the first law of the Reissner-Nordstrom black hole is recovered in the small string tension limit. Following that, we show that when the temperature of the black hole matches the temperature of the acceleration horizon, the thermodynamic partition function agrees with the semiclassical approximation of the spacetime's partition function defined through the Euclidean path integral. When the temperatures do not match, we speculate on the existence of firewalls on one or both horizons.

Jumping ahead, our C-metric first law is
\be \label{introlaw} 0 = \frac{\kap_{\rm acc}}{8\pi} \delta \Delta \ca_{\rm acc} + \frac{\kap_{\rm bh}}{8\pi} \delta \ca_{\rm bh} + \Phi \delta Q + \Delta \ell \, \delta\mu. \ee
Every term in this equation has a physical meaning, including zero! $\kap_{\rm acc}$ and $\kap_{\rm bh}$ are the surface gravities of the acceleration horizon and black hole horizon, respectively. $\Delta \ca_{\rm acc}$ is the difference in the area of the acceleration horizon between the C-metric and a related cosmic string spacetime (to be defined later). $\ca_{\rm bh}$ is the black hole horizon area. $Q$ is the physical charge of the black hole and $\Phi$ is the difference in electric potential between the acceleration horizon and the black hole horizon. $\Delta \ell$ is the difference in the ``thermodynamic length'' \cite{Appels:2017xoe} of the cosmic string between the C-metric and cosmic string spacetimes, where our thermodynamic length is the rate at which the cosmic string's worldsheet sweeps out an area in spacetime. $\mu$ is the tension of the string. Finally, the left hand side of the equation can be understood as $\delta \Delta M_{\rm boost}$ where $\Delta M_{\rm boost} = 0$ is the difference in the boost mass between the C-metric and cosmic string spacetimes. Throughout this paper we work with natural units, $c = G = \hbar = k_B = 1$.

Many individual aspects of our first law \eqref{introlaw} have appeared in previous works on accelerating black holes. Some authors have studied first laws in asymptotically locally AdS cases with no acceleration horizon \cite{Appels:2016uha, Appels:2017xoe, Anabalon:2018ydc, Anabalon}. Both of \cite{umass, hawkross} defined their notions of boost mass for the C-metric using a boost charge integral over a large hemisphere near spatial infinity, which we do as well. A quantity appearing conjugate to string tension in a first law is generally referred to as a thermodynamic length \cite{Appels:2017xoe}. In \cite{Herdeiro:2009vd, KrtousZel} a thermodynamic length essentially identical to ours, with the same worldsheet area interpretation, appears in the first law for two non-accelerating charged black holes connected by a ``strut". The work \cite{hawkross} studied the Euclidean C-metric and its action as a means of calculating the probability for a cosmic string to break via the nucleation of black holes through quantum tunneling. In \cite{Astorino} (see also \cite{acovErnst, AstReg}), the author analyzed C-metric thermodynamics with similar tools to those of this paper, but had a rather different perspective. Their first law was local to the black hole and rather than using boost time, a more general notion of time was used depending on several free functions. A first law for the C-metric was also proposed in \cite{Anabalon} by adapting methods the authors had previously used in the AdS case, but the authors warned that its thermodynamic interpretation was uncertain, and it did not incorporate the acceleration horizon as we do here. The notion of a first law applying to a particular horizon, as opposed to the patch between two horizons, is discussed in \cite{SdS} in the context of spinning de Sitter black holes. However, the mass term in those first laws comes from an integral at infinity and so they are not local in our sense.

The work we most directly build on is \cite{umass}. Using the canonical Hamiltonian formalism, its authors derived a perturbative first law based on boost time for any asymptotically locally flat spacetime with an acceleration horizon. However they did not allow the string tension to vary, their prescription for boost mass contained an ambiguity (discussed further in section \ref{sec:bkgd}), and their handling of the normalization of boost time in the case of the C-metric unnecessarily fixed a metric parameter.

\section{The C-metric}

In our coordinates, based on those of \cite{Hong:2003gx, stromzhib}, the C-metric is
\begin{equation} \label{defCmetric}
    d s_{(C)}^2 = \frac{1}{A^2(x - y)^2} \Big( G(y) d t^2 - \frac{d y^2}{G(y)} + \frac{d x^2}{G(x)} + \alpha^2 G(x) d \phi^2 \Big).
\end{equation}
It solves the Einstein equations everywhere (with a delta function stress tensor on the string), and away from the cosmic string it is an electrovacuum solution (i.e. $T_{\mu\nu} = T_{\mu\nu}^{\rm EM}$) with gauge field
\begin{equation} \label{gaugepot}
    A_\mu d x^\mu = q (y + 1) dt.
\end{equation}
Here $G(\zeta)$ is the quartic polynomial
\begin{equation}
    G(\zeta) = (1 - \zeta^2)(1 - \zeta/\zeta_1)(1 - \zeta/\zeta_2)
\end{equation}
with roots given by
\begin{equation}
    \zeta_1 = - \frac{1}{A r_-}, \hspace{0.5 cm} \zeta_2 = - \frac{1}{A r_+}, \hspace{0.5 cm} \zeta_3 = -1, \hspace{0.5 cm} \zeta_4 = 1
\end{equation}
where
\begin{equation}
    r_\pm = m \pm \sqrt{m^2 - q^2}.
\end{equation}
We require $|q| < m$ and $Ar_+ < 1$ so that the roots satisfy
\begin{equation}
    \zeta_1 < \zeta_2 < \zeta_3 < \zeta_4.
\end{equation}
Finally, we define
\begin{equation}
    \alpha = \frac{2}{|G'(1)|} = \frac{\zeta_1\zeta_2}{(1 - \zeta_1)(1 - \zeta_2)}.
\end{equation}
We can see the C-metric depends on three parameters, $m$, $q$, and $A$. Roughly speaking, $m$ is the mass of a black hole, $q$ is its charge, and $A$ is its acceleration. It should be mentioned, however, that for many calculational purposes it is simpler to instead take $\zeta_1$, $\zeta_2$, and $A$ as the three independent parameters.

The range of $x$ is $-1 \leq x \leq 1$. $\phi$ is periodically identified $\phi \sim \phi + 2 \pi$. $\alpha$ has been chosen to avoid a conical singularity at $x = 1$ between the black holes, although one remains at $x = -1$ which is the location of the cosmic string. The angular deficit around the cosmic string is
\begin{equation}\label{angulardeficit}
    \delta_{\rm def} = 2 \pi \left( 1 - \Big| \frac{G'(-1)}{G'(1)} \Big| \right).
\end{equation} The boost time $t$ can be any real number. Different ranges of $y$ correspond to different patches of the spacetime. $y \to - \infty$ is the black hole singularity, $y = \zeta_1$ is the black hole inner horizon, $y = \zeta_2$ is the black hole event horizon, $y = -1$ is the acceleration horizon, and conformal infinity is approached as $y \to x$. If $y < -1 < x$ and $x - y \to 0$, spatial infinity is reached. If $-1 < y < x$ and $y \to x$, null infinity is reached.

\begin{figure}
  \centering
  \begin{minipage}[b]{0.49\textwidth}
    \centering
        \includegraphics[width=\textwidth]{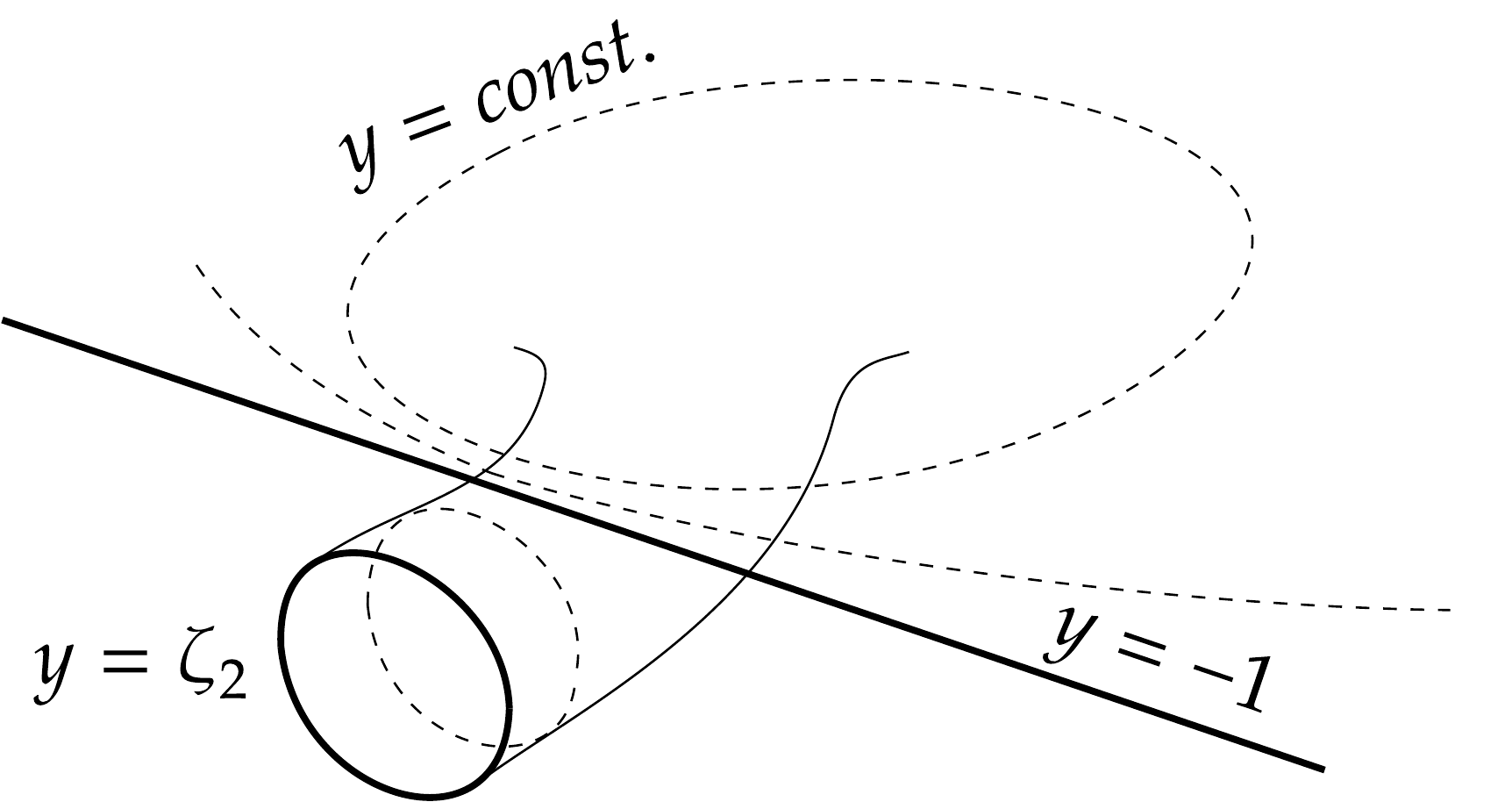}
    \caption{\label{fig:wormhole_y} Lines of constant $y$ on the $t = \rm const.$, $\phi = 0, \pi$ slice.}
  \end{minipage}
  \hfill
  \begin{minipage}[b]{0.49\textwidth}
  \centering
        \includegraphics[width=\textwidth]{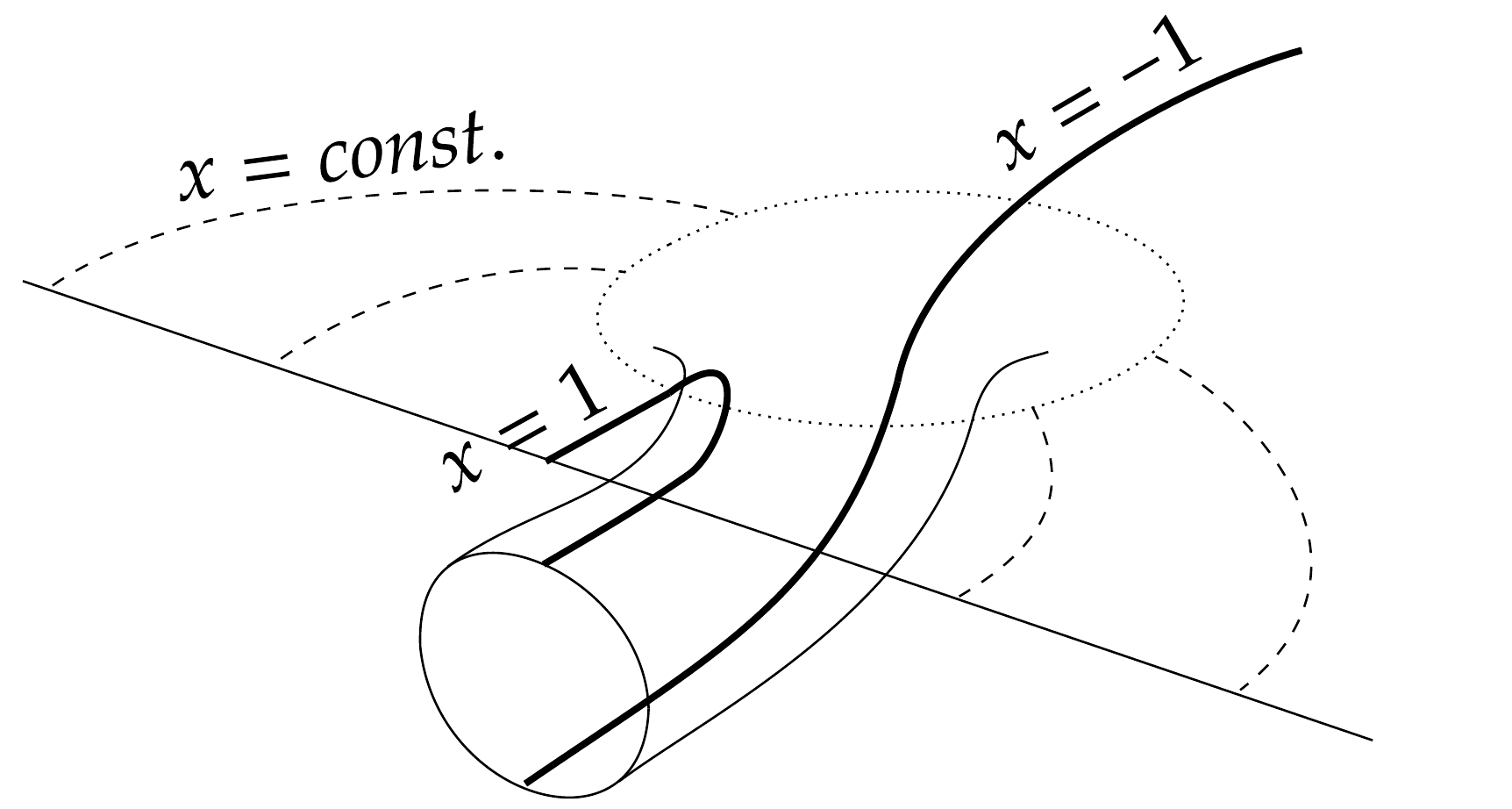}
    \caption{\label{fig:wormhole_x} Lines of constant $x$ on the $t = \rm const.$, $\phi = 0, \pi$ slice.}
  \end{minipage}
\end{figure}

The C-metric has two Killing vectors,  $\partial_\phi$ and $\partial_t$. $\partial_\phi$ generates rotations around the string while $\partial_t$ generates ``boosts'' parallel to the string. $\partial_t$ is timelike between the outer black hole horizon and acceleration horizon, but is spacelike between the acceleration horizon and null infinity, as well as between the outer and inner black hole horizons.

To get an understanding for the coordinates, in figures \ref{fig:wormhole_y} and \ref{fig:wormhole_x} we have depicted the $t = \rm const.$, $\phi = 0, \pi$ slice of the C-metric in the $\zeta_2 \leq y \leq -1$ patch. It has the geometry of a wormhole which we have drawn cut in half. We can see that on a $t = \rm const.$ slice, surfaces of constant $y$ are two-spheres. $y$ can be thought of as $-1/Ar$, where $r$ is a radial coordinate from the singularity. The variable $x$ can be thought of as $-\cos \eta$, where $\eta$ is the polar angle in bi-spherical coordinates. Note that $x$ acts as a radial coordinate on the acceleration horizon.

Different coordinates may be employed in order to study the ``worldlines'' the black holes trace out in spacetime. In \cite{gkp}, the authors do this for the uncharged C-metric and confirm that the black holes travel on roughly hyperbolic Rindler-like trajectories. 

\section{The cosmic string background spacetime} \label{sec:bkgd}

The C-metric is closely related to the flat cosmic string spacetime. This spacetime consists of nothing but a straight string in flat space whose stress-energy density sources a conical singularity. In our first law, $\Delta \ca_{\rm acc}$ and $\Delta \ell$ are defined as differences between analogous quantities computed in the C-metric and the cosmic string ``background,'' so it is worth reviewing their relationship carefully.

\begin{figure}[H]
  \centering
  \includegraphics[width=0.49\textwidth]{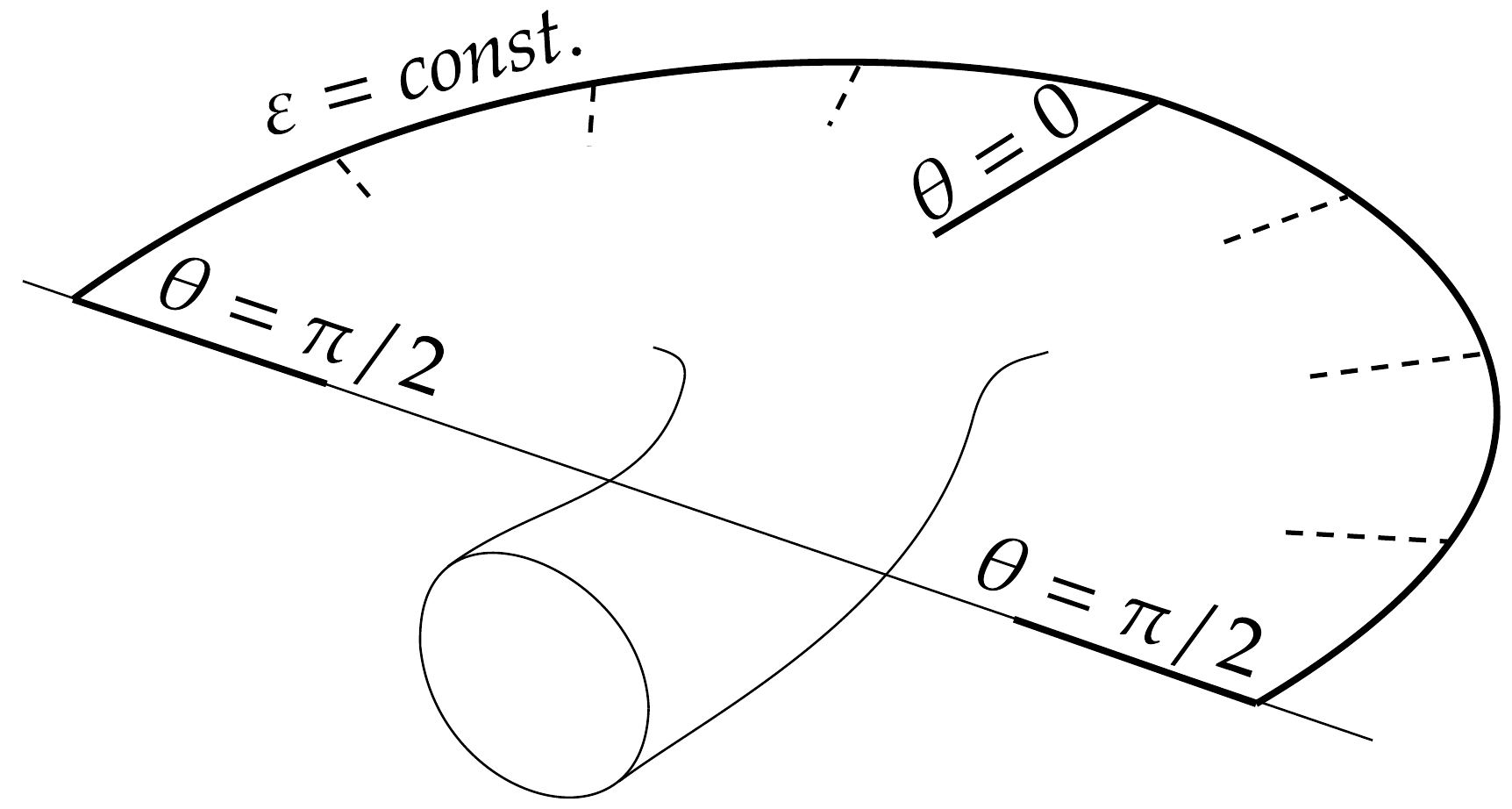}
    \caption{\label{fig:wormhole_epsilon} For $\eps \to 0$, the surface of constant $\eps$ is a large hemisphere with polar angle $\theta$. }
\end{figure}

It will be helpful to find coordinates that are naturally adapted to studying the spatial infinity of the C-metric. We can do this by parameterizing $y$ and $x$ by $\eps$ and $\theta$ as follows:
\begin{align}\label{ydef1}
    y &= -1 - \eps \cos^2 \theta, \\ \label{xdef1}
    x &= -1 + \eps \sin^2 \theta.
\end{align}
The surface of constant $\eps$ where $\eps \to 0$ corresponds to a large hemisphere approaching spatial infinity. The coordinate $\theta$ is the polar angle of this sphere. See figure \ref{fig:wormhole_epsilon}.

Using the $(t, \eps, \theta, \phi$) coordinates we will define $ds^2_{(B)}$, the exact metric of the cosmic string spacetime, to be the leading order part of the C-metric $ds^2_{(C)}$ in the small $\eps$ expansion. Here we must be sure to treat $d\eps$ as $\co(\eps)$, which renders the $d \eps d \theta$ component subleading compared to the $d\eps^2$ and $d\theta^2$ components. We find
\begin{equation}
    \begin{split}\label{dsc2def1}
    d s_{(C)}^2 & \approx \frac{4}{A^2 G'(-1) \eps} \Big( -\frac{1}{4} G'(-1)^2 \cos^2 \theta d t^2 + \frac{d \eps^2}{4\eps^2} + d \theta^2 + \sin^2 \theta \frac{G'(-1)^2}{G'(1)^2} d \phi^2 \Big) \\
    &\equiv d s^2_{(B)}.
    \end{split}
\end{equation}
One way to see that $ds^2_{(B)}$ really is the metric of a cosmic string is to substitute
\begin{align}
    R^2 &= \frac{4}{A^2 G'(-1) \eps}, \\
    \tilde{t} &= \frac{G'(-1)}{2} t,
\end{align}
at which point the metric, in $(\tilde t, R, \theta, \phi)$ coordinates, becomes
\begin{equation}
    d s^2_{(B)} = - R^2 \cos^2\theta d\tilde{t}^{\, 2}+dR^2 + R^2 d \theta^2 + \frac{G'(-1)^2}{G'(1)^2} R^2 \sin^2 \theta d \phi^2.
\end{equation}
This is the cosmic string spacetime in accelerated spherical coordinates with boost time $\tilde t$. We can see that it has the same angular deficit as the C-metric we constructed it from, given in \eqref{angulardeficit}.

Using the parameterizations of \eqref{ydef1} and \eqref{xdef1}, we have shown (by construction) that the metric components of the C-metric and cosmic string spacetimes asymptotically agree to leading order in $\eps$. However, we can do better, and doing so is crucial for later calculations. 

As discussed in \cite{hawkhor}, a proper subtraction method should match the induced metric on the boundary exactly between the spacetime and its background. The problem with \eqref{ydef1} and \eqref{xdef1} is that the induced metrics on an $\eps = \rm const.$ surface in the C-metric and cosmic string spacetimes only match at leading order in $\eps$. The first subleading order can also affect the quantities we will compute, so we need to add a correction to the coordinate transformations \eqref{ydef1} and \eqref{xdef1} that is subleading in $\eps$. An improved coordinate transformation is given by
\begin{align}\label{ydef2}
    y &= -1 - \eps \cos^2 \theta\Big( 1 - \eps \frac{(3 \cos 2 \theta -1)G''(-1)}{4 G'(-1)} \Big), \\ \label{xdef2}
    x &= -1 + \eps \sin^2 \theta \Big( 1 - \eps \frac{(3 \cos 2 \theta +1)G''(-1)}{4 G'(-1)} \Big).
\end{align}
Surprisingly, the result of matching the induced metric components on the surfaces of small constant $\eps$ ends up giving a better match of every component of the metrics $ds^2_{(C)}$ and $ds^2_{(B)}$ to one order higher in $\eps$. Namely, in these new coordinates the nonzero C-metric components satisfy
\begin{align} \nonumber
    g^{(C)}_{tt} &= g^{(B)}_{tt}(1 + \mathcal{O}(\eps^2) ), &
    g^{(C)}_{\eps \eps} &= g^{(B)}_{\eps \eps}(1 + \mathcal{O}(\eps^2) ),  \\ \label{metricmatch}
    g^{(C)}_{\theta \theta} &= g^{(B)}_{\theta \theta} (1 + \mathcal{O}(\eps^2) ), &
    g^{(C)}_{\phi \phi} &= g^{(B)}_{\phi \phi}(1 + \mathcal{O}(\eps^2) ), \\ \nonumber
    g^{(C)}_{\eps \theta} &= \co(\eps^0).
\end{align}
In an orthonormal frame these would differ from the cosmic string only at $\co(\eps^2)$. Happily, this $\co(\eps^2)$ difference will have no effect on the physical quantities we compute in the $\eps \to 0$ limit.

The calculation of boost mass in \cite{umass}, besides using ADM charges, differed from ours by relating the C-metric to the cosmic string using coordinates like our \eqref{ydef1}, \eqref{xdef1} as opposed to our \eqref{ydef2}, \eqref{xdef2}. This resulted in a different value for the boost mass.

\section{Definition of physical quantities \label{sec:defphys}}

In this section we provide formulas for all physical quantities appearing in our first law \eqref{introlaw}. We note that every integration performed herein occurs on a $t = \rm const.$ slice. We begin with the physical charge and black hole area,
\begin{equation}
    Q = \frac{1}{4 \pi} \int \star F = \frac{1}{4 \pi} \int \sqrt{-g^{(C)}}F^{t y} dx d \phi\Bigg\rvert_{y = \rm const.} = \alpha q = \frac{\sqrt{\zeta_1\zeta_2}}{A(1 - \zeta_1)(1 - \zeta_2)},
\end{equation}
\begin{equation}
    \ca_{\rm bh} = \int \sqrt{g^{(C)}_{xx} g^{(C)}_{\phi \phi}} d x d \phi \Bigg\rvert_{y = \zeta_2 } = \frac{4 \pi \zeta_1 \zeta _2}{A^2 (\zeta _1-1) (\zeta _2-1)^2 (\zeta _2+1)}.
\end{equation}
The string tension $\mu$ is related to the angular deficit $\delta_{\rm def}$ (given in \eqref{angulardeficit}) by $\delta_{\rm def} = 8 \pi G \mu$. Here we set $G = 1$, giving
\begin{equation} \label{mudef}
    \mu = \alpha A m = \frac{-\zeta_1 - \zeta_2}{2 (1 - \zeta_1) (1 - \zeta_2)}.
\end{equation}

To compute the change in the area of the acceleration horizon, $\Delta \ca_{\rm acc}$, we must compute its area in the C-metric, $\ca_{\rm acc}^{(C)}$, and its area in the background, $\ca_{\rm acc}^{(B)}$. We find
\begin{equation}\label{AaccC}
    \ca_{\rm acc}^{(C)} = \int \sqrt{g^{(C)}_{xx} g^{(C)}_{\phi \phi}} dx d \phi \Bigg\rvert_{y = -1 } =  \frac{2 \pi \alpha}{A^2}\int^{1}_{x_{\rm min}} \frac{dx}{(x+1)^2} = \frac{2 \pi \alpha}{A^2} \Big( \frac{1}{x_{\rm min} + 1} - \frac{1}{2}\Big),
\end{equation}
\begin{equation}\label{AaccB}
    \ca^{(B)}_{\rm acc} = \int \sqrt{g^{(B)}_{\eps \eps} g^{(B)}_{\phi \phi}} d \eps d \phi \Bigg\rvert_{\theta = \pi/2 } = \frac{2 \pi \alpha}{A^2} \int^{\infty}_{\eps_{\rm min}} \frac{d \eps}{\eps^2} = \frac{2 \pi \alpha}{A^2 \eps_{\rm min}}.
\end{equation}
As we integrate over the whole acceleration horizon, $x_{\rm min} \to -1$ and $\eps_{\rm min} \to 0$. Using \eqref{xdef2} at $\theta = \pi/2$, we have
\begin{equation}
    x_{\rm min} = -1 + \eps_{\rm min} \Big( 1 + \eps_{\rm min}\frac{G''(-1)}{2 G'(-1)} \Big).
\end{equation}
In the $\eps_{\rm min} \to 0$ limit, we get
\begin{equation}
    \Delta \ca_{\rm acc} = \ca_{\rm acc}^{(C)} - \ca^{(B)}_{\rm acc} = \frac{2 \pi  \zeta_1 \zeta_2 (\zeta_1+\zeta_2+2)}{A^2 \left(\zeta_1^2-1\right) \left(\zeta_2^2-1\right)}.
\end{equation}
The remaining quantities $\kap_{\rm acc}$, $\kap_{\rm bh}$, $\Phi$, and $\Delta \ell$ all depend on the normalization of our time coordinate. As it is not immediately clear what the most natural normalization of $t$ is, we will define the Killing vector
\begin{equation} \label{xidef}
    \xi = \xi^\mu \partial_\mu = N \partial_t
\end{equation}
with some constant normalization $N$ which we leave unspecified for now. Eventually we will see that different choices of $N$ are suitable in different contexts.

We now compute the surface gravities $\kap_{\rm acc}$ and $\kap_{\rm bh}$. It is interesting to note that $\xi$ becomes null on both the acceleration horizon and the black hole horizon, meaning we can use the same vector $\xi$ to calculate the surface gravity of both horizons. Recall that surface gravity is given by
\begin{equation}
    \kap= \sqrt{\nabla_\mu V \nabla^\mu V}, \hspace{0.5 cm} \text{where} \hspace{0.5 cm} V =\sqrt{-\xi^\mu \xi_\mu},
\end{equation}
evaluated on the horizon in question. This gives
\begin{equation}
    \kap_{\rm acc} = \frac{N}{2} G'(-1) = N \left(1 + \zeta_1^{-1}\right) \left(1 + \zeta_2^{-1}\right),
\end{equation}
\begin{equation}
    \kap_{\rm bh} = \frac{N}{2} |G'(\zeta_2)| = N \frac{\left(\zeta_2^2-1\right) (\zeta_2-\zeta_1)}{2 \zeta_1 \zeta_2}.
\end{equation}

Next we define the values of the electric potential at the black hole horizon, the acceleration horizon, and spatial infinity:
\begin{align}
    \Phi_{\rm bh} & = (\xi^\mu A_\mu)\Big\rvert_{y = \zeta_2} = Nq(\zeta_2+1) = N \frac{\zeta_2 + 1}{A \sqrt{\zeta_1 \zeta_2}}, \\
    \Phi_{\rm acc} & = \Phi_\infty = (\xi^\mu A_\mu)\Big\rvert_{y = -1} = 0.
\end{align}
Despite the vanishing of the latter two with our choice of gauge potential in \eqref{gaugepot}, we find it instructive to keep them explicit in some contexts. We also define the gauge-invariant quantity
\be \Phi \equiv \Phi_{\rm acc} - \Phi_{\rm bh}. \ee

Finally, we must discuss the physical interpretation of the ``thermodynamic length" \cite{Appels:2017xoe}. As the string evolves along the timelike Killing vector $\xi$, it sweeps out a worldsheet in spacetime. We take our thermodynamic length as the rate at which the area of the worldsheet increases as the Killing time progresses. $\Delta \ell$ is the difference between the thermodynamic length of the C-metric and the cosmic string background.

We begin by computing the thermodynamic lengths of the C-metric and cosmic string background separately. We find
\begin{equation}
    \ell^{(C)} = N \int^{y_{\rm max}}_{\zeta_2} dy \sqrt{-g^{(C)}_{tt} g^{(C)}_{yy}}\Bigg\rvert_{x = -1} = \frac{N}{A^2} \Big( \frac{1}{1 + \zeta_2} - \frac{1}{1 + y_{\rm max}} \Big),
\end{equation}
\begin{equation}
    \ell^{(B)} = N \int^{\infty}_{\eps_{\rm min}} d\eps \sqrt{-g^{(B)}_{tt} g^{(B)}_{\eps \eps}}\Bigg\rvert_{\theta = 0} = \frac{N}{A^2 \eps_{\rm min}}.
\end{equation}
Using \eqref{ydef2} at $\theta = 0$, we find
\begin{equation}
    y_{\rm max} = -1 - \eps_{\rm min} \Big( 1 - \eps_{\rm min} \frac{G''(-1)}{2 G'(-1)} \Big).
\end{equation}
In the $\eps_{\rm min} \to 0$ limit, we then find
\begin{equation}
    \Delta \ell = \ell^{(C)} - \ell^{(B)} = -\frac{N}{2 A^2} \frac{3 + \zeta_1}{1 + \zeta_1}.
\end{equation}
This quantity may seem somewhat obscure, but it is closely related to the Nambu-Goto action of the string and has appeared in a similar context in \cite{Herdeiro:2009vd, KrtousZel}.

We have now given a physical interpretation to every quantity in \eqref{introlaw} and expressed them in terms of the three independent parameters $\zeta_1$, $\zeta_2$, and $A$ (and the arbitrary normalization $N$). It can be explicitly verified that \eqref{introlaw} holds by writing everything in terms of $\zeta_1, \zeta_2, A$ and applying the chain rule when computing the variations $\delta \Delta \ca_{\rm acc}, \delta \ca_{\rm bh}, \delta Q, \delta \mu$.

These quantities also satisfy the Smarr relation
\begin{equation}\label{smarr}
    0 = \frac{\kap_{\rm acc}}{8 \pi} \Delta \ca_{\rm acc} + \frac{\kap_{\rm bh}}{8 \pi} \ca_{\rm bh} + \frac{1}{2}\Phi Q
\end{equation}
which essentially follows from \eqref{introlaw}, Euler's theorem for homogeneous functions, and the scaling dimensions of the quantities involved. $\mu$ does not appear as it is dimensionless.

\section{Deriving the first law(s) with the covariant phase space formalism}

Noetherian methods in the covariant phase space formalism offer an elegant derivation of the first law for Kerr-Newman black holes \cite{wald1993black, bhmech}. Given a Killing vector $\xi$ and field variations $\delta g_{\mu \nu}$ and $\delta A_\mu$ one can construct a 2-form $k_\xi$ which is closed, satisfying $d k_\xi = 0$, away from matter sources. The integral of $k_\xi$ over a 2-surface gives the variation of the enclosed $\xi$-charge. This quantity is invariant under continuous deformations of the surface that do not pass through matter. It should be noted that this charge variation is not necessarily integrable --- i.e. it is not necessarily the variation of a well-defined finite charge. Integrability must be checked separately. When $\xi$ is a time translation, one refers to the $\xi$-charge as the energy (or mass). Similarly when $\xi$ is a boost we will refer to the $\xi$-charge as the ``boost mass" \cite{umass}.

In Einstein-Maxwell gravity this 2-form has two contributions, one ``gravitational'' and one ``electromagnetic.'' Explicitly,
\begin{equation} \label{kdeffirst}
    k_\xi = k^{\rm grav}_\xi + k^{\rm em}_\xi 
\end{equation}
where
\begin{align}
    k^{\rm grav}_\xi &= - \delta K_\xi^{\rm grav}  + K_{\delta \xi}^{\rm grav} - i_\xi \Theta^{\rm grav}, \\
    K_\xi^{\rm grav}  &= \frac{1}{16 \pi} \cdot \frac{1}{4} \eps_{\mu \nu \rho \sigma} (\nabla^\mu \xi^\nu - \nabla^\nu \xi^\mu) dx^\rho \wedge dx^\sigma, \\
    i_\xi \Theta^{\rm grav} &= - \frac{1}{16 \pi} \cdot \frac{1}{2} \eps_{\mu\nu\rho\sigma} \xi^\nu (\nabla_\alpha \delta g^{\alpha \mu} - \nabla^\mu (g_{\alpha \beta} \delta g^{\alpha \beta} )) dx^\rho \wedge dx^\sigma,
\end{align}
and
\begin{align}
    k^{\rm em}_\xi &= - \delta K^{\rm em}_\xi + K^{\rm em}_{\delta \xi} - i_\xi \Theta^{\rm em}, \\
    K^{\rm em}_\xi &= \frac{1}{16 \pi} \cdot \frac{1}{4}\eps_{\mu \nu \rho \sigma} (F^{\mu \nu} \xi^\alpha A_\alpha ) dx^\rho \wedge dx^\sigma, \\
    i_\xi \Theta^{\rm em} &= \frac{1}{16 \pi} \cdot \frac{1}{2} \eps_{\mu\nu\rho\sigma} \xi^\nu F^{\alpha \mu} \delta A_\alpha dx^\rho \wedge dx^\sigma. \label{kdeflast}
\end{align}
Here $\eps_{\mu \nu \rho \sigma}$ is the totally antisymmetric tensor with $\eps_{0123} = \sqrt{-g}$. The formalism works for generic variations $\delta g_{\mu \nu}, \delta A_\mu$, but often one works within a parametrized family of solutions, for example the Kerr-Newman or C-metric families.

The derivation of the first law for Reissner-Nordstrom black holes using covariant phase space methods is instructive, so we briefly outline it here. $\xi$ is chosen to be the horizon generator of the black hole, normalized so that $\xi^2 = -1$ at spatial infinity. The physical motivation for this normalization is that $\xi$ coincides with sense of time (i.e. four-velocity) of a distant static observer. In the standard Reissner-Nordstrom coordinates, $\xi = \partial_t$.\footnote{In this discussion on Reissner-Nordstrom black holes we temporarily redefine $\xi$ and other quantities.} For our surface, we take the union of a large two-sphere at spatial infinity and the black hole bifurcation two-sphere with opposite orientations. As this surface is contractible, the integral of $k_\xi$ over it must vanish. When $\int k^{\rm grav}_\xi$ is evaluated on the horizon it reduces to $\frac{\kap_{\rm bh}}{8\pi} \delta \ca_{\rm bh}$ and when $\int k_\xi^{\rm em}$ is evaluated on the horizon it gives $-\Phi_{\rm bh} \delta Q$. The variation in energy of the spacetime is given by $\int k_\xi$ evaluated at spatial infinity. When we use the standard gauge where $\Phi_{\infty} = 0$, this integral is equal to $\delta M$ where $M$ is the usual mass of the black hole. From the contractibility of the total surface, these quantities sum to zero, giving the first law for Reissner-Nordstrom black holes.

One might call this the ``global'' version of the first law, associated with the full spacetime, as opposed to a version ``local'' to the horizon. Local first laws, described herein, are appropriate to use in contexts where there are multiple horizons or matter present. For pure Reissner-Nordstrom the difference between the global and local first law is immaterial because the sphere on the horizon can be continuously deformed to the sphere at infinity without passing through matter. But if we take a charged black hole with matter outside it (such that we still have a timelike Killing vector) then this is no longer the case. The difference in the surface integral on the horizon and at spatial infinity would be given by integrals on ``bubbles'' surrounding the parcels of matter. The surface integral at spatial infinity therefore gives the energy variation of the black hole plus that of the external matter. In contrast, the surface integral on the black hole itself gives only the black hole's energy variation and can be written in terms of quantities local to the black hole such as its area and charge.

All these general comments apply in particular to the C-metric. It has the string as a matter source as well as two horizons. Our main first law \eqref{introlaw} is of the global type, but there is also a local first law for each horizon. The local first law of the black hole is precisely the one discussed in \cite{Astorino}. As we will see later, in the small tension limit the results of Reissner-Nordstrom are most easily reproduced using not our global first law but rather the local law of the black hole.

\begin{figure}
  \centering
  \includegraphics[width=0.8\textwidth]{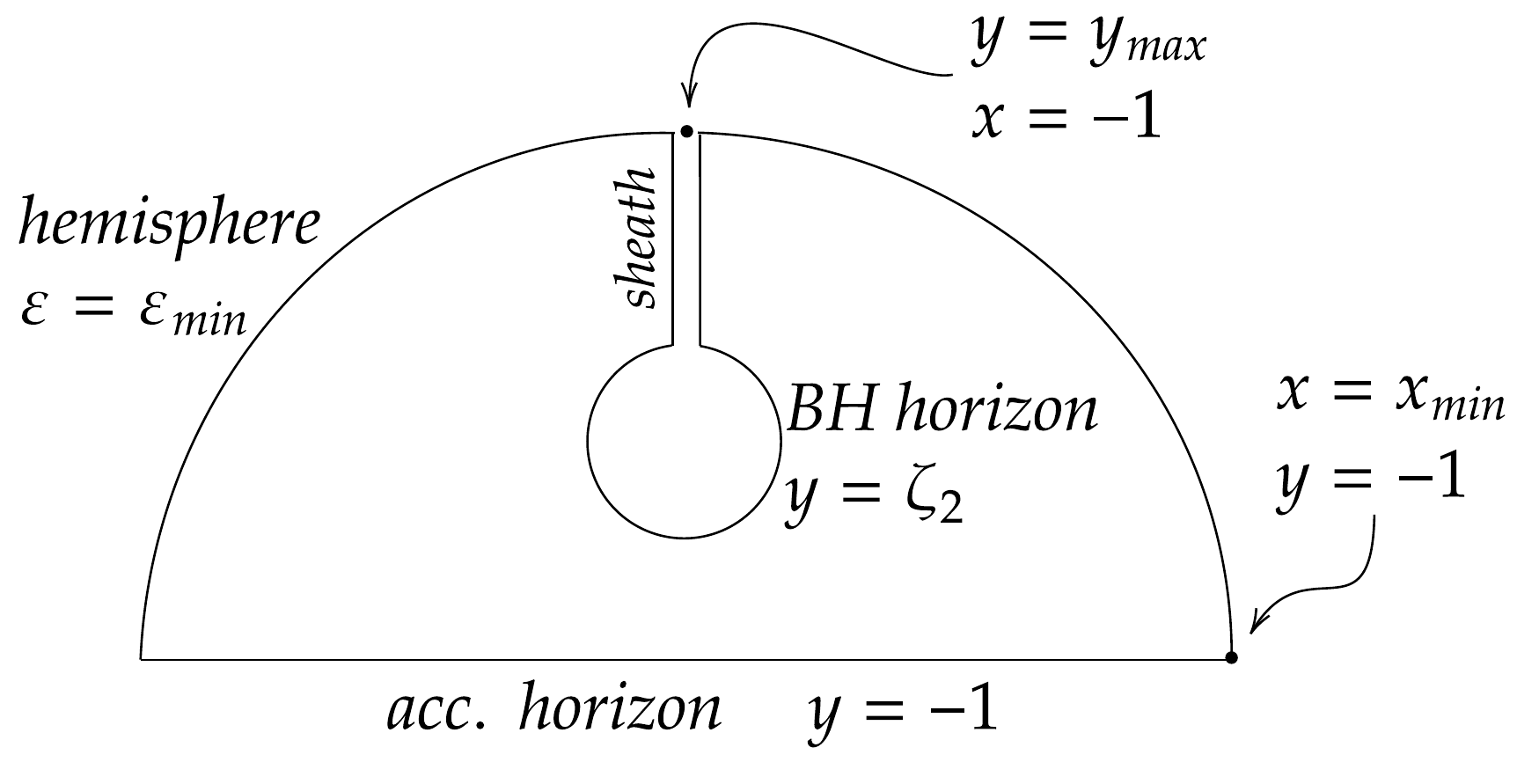}
    \caption{\label{fig:surface} This surface can be contracted to a point without passing through the string, which is located in the thin sheath. In this diagram $\phi$ has been suppressed.}
\end{figure}

We now turn to the actual derivation of the C-metric's first laws. We take our system to be the $\zeta_2 \leq y \leq -1$ patch, which is bounded by the acceleration horizon, the black hole horizon, and spatial infinity. We can parametrize our phase space of C-metrics with the three numbers $\zeta_1$, $\zeta_2$, and $A$. For our Killing vector we take $\xi = N \p_t$ as in \eqref{xidef}, which is the generator of both horizons. For now we leave the normalization constant $N$ to be any function on phase space, $N = N(\zeta_1, \zeta_2, A)$. In order to calculate $k_\xi$ we must specify what our metric and gauge field variations are, using $\delta$ for the differential on phase space. For example
\begin{equation}
    \delta g_{\mu \nu} = \frac{\p g_{\mu \nu}}{\p \zeta_1} \delta \zeta_1 + \frac{\p g_{\mu \nu}}{\p \zeta_2} \delta \zeta_2 + \frac{\p g_{\mu \nu}}{\p A} \delta A.
\end{equation}
For the global first law, we must construct a two-dimensional surface that is contractible to a point without passing through the string. This surface is depicted in figure \ref{fig:surface}. Embedded in a constant $t$ slice, the surface begins by surrounding most of the black hole horizon before connecting to a thin sheath which encases the string. The sheath then connects to a large hemisphere at spatial infinity, which is capped off by the acceleration horizon. This large hemisphere is the surface of constant $\eps = \eps_{\rm min}$ where $\eps_{\rm min} \to 0$.

We break the integral over this surface into four parts and evaluate them using \eqref{kdeffirst} -- \eqref{kdeflast}. The explicit results in terms of $\zeta_1$, $\zeta_2$, $A$, $\delta \zeta_1$, $\delta \zeta_2$, and $\delta A$ are unenlightening, but can be expressed more briefly in terms of quantities defined in section \ref{sec:defphys}. We find
\begin{align}
    \label{onbh} \slashed{\delta} M_{\rm bh} &\equiv \int_{\rm bh} k_\xi = \frac{\kap_{\rm bh}}{8\pi} \delta \ca_{\rm bh} - \Phi_{\rm bh} \delta Q, \\
    \label{onacc} \slashed{\delta} M^{(C)}_{\rm acc} &\equiv \int_{\rm acc} k_\xi = \frac{\kap_{\rm acc}}{8\pi} \delta \ca_{\rm acc}^{(C)} + \Phi_{\rm acc} \delta Q, \\
    \label{onhemi} \slashed{\delta} M^{(C)}_{\rm hemi} &\equiv \int_{\rm hemi} k_\xi = -\frac{\kap_{\rm acc}}{8\pi} \delta \ca^{(B)}_{\rm acc} - \ell^{(B)} \delta \mu, \\
    \label{onsheath} \slashed{\delta} M_{\rm sheath}^{(C)} &\equiv \int_{\rm sheath} k_\xi = \ell^{(C)} \delta \mu.
\end{align}
Each quantity $\slashed{\delta} M$ is meaningful on its own as a variation in boost mass. The local first law of the black hole is \eqref{onbh}, as in \cite{Astorino}. Note that the quantities $\slashed{\delta} M_{\rm bh}$, $\kap_{\rm bh}$, $\ca_{\rm bh}$, $\Phi_{\rm bh}$, and $Q$ can all be computed locally on the black hole horizon. Similarly we identify \eqref{onacc} as the local first law of the acceleration horizon and note that all of its quantities can be computed locally as well. We identify $-\slashed{\delta} M^{(C)}_{\rm hemi}$ as the boost mass of the system (with a sign for orientation) since it is the charge evaluated at spatial infinity, in analogy with Reissner-Nordstrom. Both \cite{hawkross} and \cite{umass} define their boost masses using this same surface. Lastly $\slashed{\delta} M^{(C)}_{\rm sheath}$ is simply the boost mass contribution of the string. As these quantities sum to zero due to the contractibility of the surface, it appears we are ready to derive our first law. We have
\be \begin{aligned} \label{Cfirstlaw}
    - \slashed{\delta} M^{(C)}_{\rm hemi} &= \slashed{\delta} M_{\rm bh} + \slashed{\delta} M_{\rm acc}^{(C)} + \slashed{\delta} M_{\rm sheath}^{(C)} \\
    &= \frac{\kap_{\rm bh}}{8 \pi} \delta \ca_{\rm bh} + \frac{\kap_{\rm acc}}{8 \pi} \delta \ca^{(C)}_{\rm acc} + \Phi \delta Q + \ell^{(C)} \delta \mu.
\end{aligned} \ee
The left hand side is the variation of the system's boost mass, and the right hand side has the usual form of a first law, being a sum of products of conjugate variables. However, our expression as written is not fully satisfactory since the ``bare" quantities $\slashed{\delta} M^{(C)}_{\rm hemi}$, $\ca^{(C)}_{\rm acc}$, and $\ell^{(C)}$ all diverge as $\eps_{\rm min} \to 0$. A little rearrangement by subtracting $\frac{\kap_{\rm acc}}{8\pi} \delta \ca^{(B)}_{\rm acc}$ and $\ell^{(B)} \delta \mu$ from both sides would solve this at the level of equations, reproducing \eqref{introlaw}. However, it is preferable to regularize terms in a principled manner so that each term in the equation retains a physical interpretation and in particular that the left hand side can still be considered a boost mass. The natural way to regularize is by comparing each term with the corresponding quantity in the cosmic string background. To that end, we construct a surface analogous to the one in figure \ref{fig:surface} in the cosmic string spacetime. The main difference is that, as there is no black hole in this spacetime, the sheath connects directly to the acceleration horizon. We may then evaluate the surface charges just as before, finding
\begin{align}
    \slashed{\delta} M^{(B)}_{\rm acc} &\equiv \int_{\rm acc} k_\xi =  \frac{\kap_{\rm acc}}{8 \pi} \delta \ca^{(B)}_{\rm acc}, \\
    \slashed{\delta} M^{(B)}_{\rm hemi} &\equiv \int_{\rm hemi} k_\xi = -\frac{\kap_{\rm acc}}{8\pi} \delta \ca^{(B)}_{\rm acc} - \ell^{(B)} \delta \mu, \\
    \slashed{\delta} M^{(B)}_{\rm sheath} &\equiv \int_{\rm sheath} k_\xi = \ell^{(B)} \delta \mu.
\end{align}
As this analogous surface is also contractible, we have the relation
\begin{equation} \label{Bfirstlaw}
    -\slashed{\delta}M_{\rm hemi}^{(B)} = \slashed{\delta}M_{\rm acc}^{(B)} + \slashed{\delta}M_{\rm sheath}^{(B)}.
\end{equation}
We now define our regularized first law by subtracting \eqref{Bfirstlaw} from \eqref{Cfirstlaw} term by term. Defining
\begin{align}
    \label{reghemi} \slashed{\delta}\Delta M_{\rm hemi} &\equiv \slashed{\delta} M_{\rm hemi}^{(C)} - \slashed{\delta} M_{\rm hemi}^{(B)} = 0, \\
    \label{regacc} \slashed{\delta} \Delta M_{\rm acc} &\equiv \slashed{\delta} M_{\rm acc}^{(C)} - \slashed{\delta} M_{\rm acc}^{(B)} = \frac{\kap_{\rm acc}}{8 \pi} \delta \Delta \ca_{\rm acc} + \Phi_{\rm acc} \delta Q, \\
    \label{regsheath} \slashed{\delta} \Delta M_{\rm sheath} &\equiv \slashed{\delta} M_{\rm sheath}^{(C)} - \slashed{\delta} M_{\rm sheath}^{(B)} = \Delta \ell \, \delta \mu,
\end{align}
the subtracted first law reads
\be \begin{aligned} \label{thefirst}
    -\slashed{\delta}\Delta M_{\rm hemi} &=  \slashed{\delta} \Delta M_{\rm acc} + \slashed{\delta} M_{\rm bh}+ \slashed{\delta} \Delta M_{\rm sheath} \\
    0 &= \frac{\kap_{\rm acc}}{8\pi} \delta \Delta \ca_{\rm acc} + \frac{\kap_{\rm bh}}{8\pi} \delta \ca_{\rm bh} + \Phi \delta Q + \Delta \ell \, \delta \mu.
\end{aligned} \ee
This completes the derivation of our regularized first law. We may define the left hand side as the regularized boost mass of the spacetime
\begin{equation}
    \delta \Delta M_{\rm boost} \equiv - \slashed{\delta} \Delta M_{\rm hemi} = 0.
\end{equation}
Note that $\delta \Delta M_{\rm boost} = 0$ is trivially integrable, yielding $\Delta M_{\rm boost} = 0$. This vanishing of the boost mass is consistent with \cite{hawkross}. It follows directly from the fact that the C-metric and cosmic string spacetimes match at spatial infinity to subleading order in $\eps$.

We may now inspect our first law and try to assign a temperature to the spacetime using the formula $T = \kap/2 \pi$. However, we see the appearance of two possible temperatures, one for the acceleration horizon and one for the black hole horizon. The proper way to assign temperature when there are multiple horizons is still an open question \cite{SdS}. However when the two temperatures are equal, a direct thermodynamic interpretation is possible. This is discussed in section \ref{temp matched}. In section \ref{firewall} we briefly speculate on how to handle temperature with distinct surface gravities.

One may wonder if $\slashed{\delta} M_{\rm bh}$ and $\slashed{\delta} \Delta M_{\rm acc}$ can be integrated, as doing so would allow a finite boost mass to be ascribed to the acceleration horizon and black hole horizon. We can use gauge transformations and choices of $N$, the normalization of our horizon generator $\xi$, to do this.

We begin with the acceleration horizon,
\begin{equation}
     \slashed{\delta} \Delta M_{\rm acc} = \frac{\kap_{\rm acc}}{8 \pi} \delta \Delta \ca_{\rm acc} + \Phi_{\rm acc} \delta Q.
\end{equation}
Similar to the case of Reissner-Nordstrom, it is natural to choose the gauge and normalization which would naturally be used by a distant static observer. That is, use the gauge \eqref{gaugepot}, for which $\Phi_{\rm acc} = \Phi_\infty = 0$, and normalize time to agree with asymptotic Rindler time, which means $\kap_{\rm acc} = 1$. In this case $\slashed{\delta} \Delta M_{\rm acc}$ is trivially integrable as
\begin{equation}
    \slashed{\delta} \Delta M_{\rm acc} = \delta ( \frac{1}{8 \pi} \Delta \ca_{\rm acc}).
\end{equation}
While of questionable physical content, we believe this is the statement that most closely comprises a first law of the acceleration horizon. Note that since the regularized hemisphere charge vanishes, $\Delta M_{\rm acc}$ also equals the boost mass of the black hole plus string subsystem. Since the black hole and string are coupled we cannot generally write it as a sum of two separate boost masses.

We now turn to the black hole horizon, with
\begin{equation}
     \slashed{\delta}M_{\rm bh} = \frac{\kap_{\rm bh}}{8 \pi} \delta \ca_{\rm bh} - \Phi_{\rm bh} \delta Q.
\end{equation}
The perspective of a distant static observer is not integrable here, and the somewhat trivial choice $\kap_{\rm bh} = 1$, $\Phi_{\rm bh} = 0$ just reproduces the area $\ca_{\rm bh}$. There are no other clear natural choices of normalization $N$ or gauge. If we allow ourselves complete freedom in choosing them, we have the power to not only make $\slashed{\delta}M_{\rm bh}$ integrable, but to make $M_{\rm bh}$ \textit{any} function of $\ca_{\rm bh}$ and $Q$. Explicitly, given any $M_{\rm bh}(\ca_{\rm bh}, Q)$, all we need to do is choose an $N$ such that $\kap_{\rm bh} = 8\pi \frac{\p M_{\rm bh}}{\p \ca_{\rm bh}}$ and choose the gauge so that $\Phi_{\rm bh} = -\frac{\p M_{\rm bh}}{\p Q}$.

If we want our mass to be consistent with the Reissner-Nordstrom black hole, then we can choose $M_{\rm bh}$ to satisfy the Christodolou-Ruffini formula
\be M_{\rm bh}(\ca_{\rm bh}, Q) = \frac{\sqrt{\ca_{\rm bh}/\pi}}{4} \left( 1 + \frac{4\pi Q^2}{\ca_{\rm bh}} \right). \ee
This is essentially the point of view taken by \cite{Astorino} (see also \cite{acovErnst, AstReg}). In our view this result is something one has chosen rather than found unless the choice of normalization and gauge were physically motivated. However, as we will see, when the string tension is small there is a natural choice that recovers the Reissner-Nordstrom mass.

\section{The small string tension limit}

Recall the charged C-metric is specified by three parameters. Two can be taken dimensionless, such as $\zeta_1, \zeta_2$, but the third parameter, such as $A$, must be dimensionful. Let us now take the two dimensionless parameters to be $\mu$ and $\rho$, where $\rho$ is the ``charge to mass ratio"
\be \rho \equiv q/m. \ee
It is possible to express $\zeta_1$ and $\zeta_2$ as functions of $\mu$ and $\rho$.

In this section we explore the limit of small $\mu$ when $\rho$ and $A$ remain finite. Since $\mu = \alpha A m$ and $\alpha = 1 + \co(\mu)$, small $\mu$ is equivalent to
\be A m \ll 1. \ee
Physically this means that the black hole's length scale $m$ is small compared to the acceleration length scale $A^{-1}$. Therefore the black hole can be understood as a pointlike Rindler particle from the perspective of a distant observer \cite{kinwalk}. To an observer close to the black hole, however, the black hole is barely accelerating and resembles a Reissner-Nordstrom black hole at rest. These two notions of scale, one far from the black hole and one intermediate, lead to different notions of the mass of the black hole.\footnote{
In fact, when $\mu \to 0$, the C-metric \eqref{defCmetric} can be shown to reduce to the Minkowski metric and the Reissner-Nordstrom metric using different limiting procedures. If we fix $A, \rho$ and send $\mu \to 0$ the resulting metric describes Minkowski space. If we fix $m, \rho$ and switch to $t_{RN} \equiv t/A$, and send $\mu \to 0$, the resulting metric describes a Reissner-Nordstrom black hole of mass $m$ and charge $q$.}

\begin{figure}
  \centering
  \includegraphics[width=.78\textwidth]{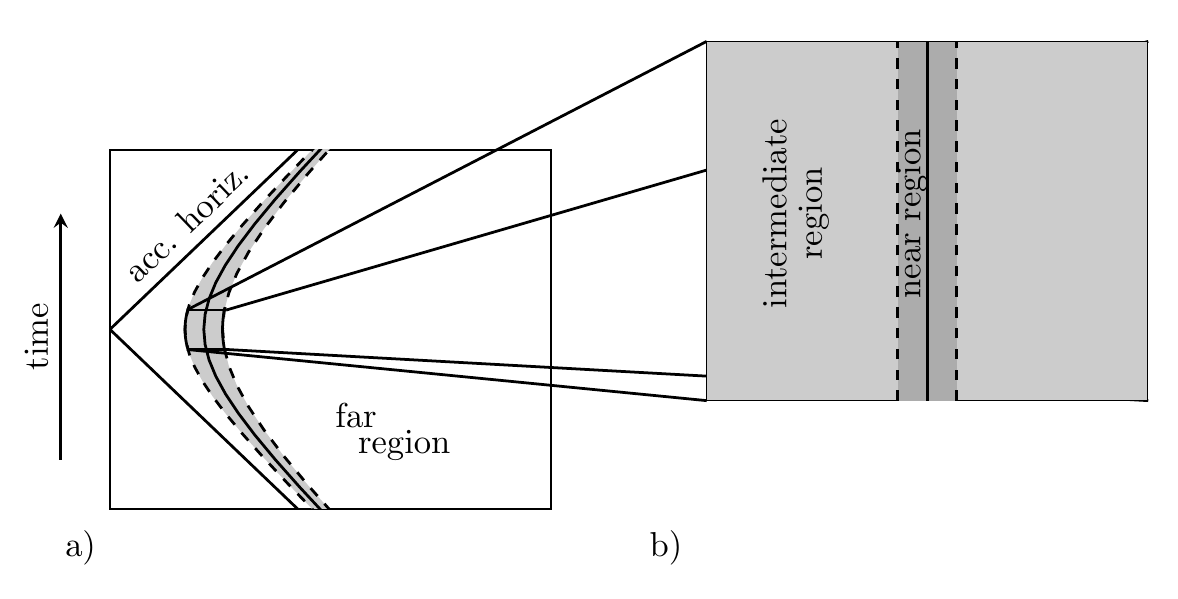}
    \caption{\label{fig:scales} \textbf{a)} In the ``far region" outside the dotted lines the black hole appears pointlike, tracing out a hyperbolic trajectory in time. \textbf{b)} In the ``intermediate region" outside the dotted lines the black hole still appears pointlike, but has negligible acceleration at this scale.}
\end{figure}

With our choice of gauge in \eqref{gaugepot}, where $\Phi_{\rm acc} = 0$, the local first law of the black hole reduces at small $\mu$ to
\be \slashed{\delta} M_{\rm bh} = \frac{\kap_{\rm bh}}{8 \pi} \delta \Delta \ca_{\rm bh} - \Phi_{\rm bh} \delta Q \approx N \left( -\frac{\mu}{A^3} \delta A + \frac{1}{A^2} \delta \mu \right). \ee
We will use this equation to study both the point particle and Reissner-Nordstrom limits. First consider the perspective of a distant observer where the black hole appears pointlike. In Minkowski space with coordinates $(X^0, X^1, X^2, X^3)$ the canonical boost generator is
\be K = X^3 \p_0 + X^0 \p_3 \ee
and a Rindler point particle of mass $m_{\rm pt}$ and acceleration $A_{\rm pt}$ has boost mass
\be M_{\rm pt} = \int d^3x \, K^\mu T_{\mu 0} = \frac{m_{\rm pt}}{A_{\rm pt}}. \ee
To compare with the black hole we start by normalizing $\xi$ to agree with $K$ asymptotically. This means $\kap_{\rm acc} = 1$, or $N \approx 1$. Now at leading order in $\mu$ we have
\be \slashed{\delta} M_{\rm bh} \approx -\frac{\mu}{A^3} \delta A + \frac{1}{A^2} \delta \mu. \ee
Although $\slashed{\delta} M_{\rm bh}$ is not integrable, $M_{\rm bh}$ can be perturbatively defined around $\mu = 0$. For any finite $A$, we know $\mu = 0$ corresponds to Minkowski space and so the (absent) black hole's mass vanishes. If we increase $\mu$ slightly from zero then the boost mass is
\be M_{\rm bh} \approx \frac{\mu}{A^2} \approx \frac{m}{A}. \ee
This agrees with the boost mass of the point particle. Note we can reproduce this answer using $\Delta M_{\rm acc}$, which equals the combined boost mass of the black hole and string. One can check that for small $\mu$,
\be \Delta M_{\rm acc} \approx \frac{m}{A} + \mu \Delta \ell\ee
where the first term is the boost mass of the black hole and the second term is the boost mass of the string.

Now we turn to the perspective where we have zoomed in on the black hole. This region is characterized by $y \ll -1$, meaning the observer is much closer to the black hole than the acceleration horizon. However, if we also demand $y / \zeta_2 \ll 1$ then the observer will also be much further away from the black hole than the length scale $m$. The region described by both limits is called the ``intermediate region,'' and can be thought of as the analogue of the region far from the black hole in the pure Reissner-Nordstrom spacetime. As usual we normalize by matching $\xi$ with the four-velocity of our static observer, which here requires $N \approx A$. This physically motivated choice of $N$ yields
\be \slashed{\delta} M_{\rm bh} \approx -\frac{\mu}{A^2} \delta A + \frac{1}{A} \delta \mu \approx \delta  \left( \frac{\mu}{A} \right) \approx \delta m, \ee
\be \implies M_{\rm bh} \approx m \ee
which of course agrees with Reissner-Nordstrom. The black hole surface gravity, with $N = A$ in the small $\mu$ limit, equals
\be \kap_{\rm bh} \approx \frac{\sqrt{m^2 - q^2}}{(m + \sqrt{m^2 - q^2})^2}, \ee
which also agrees with Reissner-Nordstrom. In fact, $\Phi$ and $\ca_{\rm bh}$ also reduce to their Reissner-Nordstrom values in the small $\mu$ limit, meaning the whole first law is reproduced.

\section{The Euclidean C-metric} \label{Euclidean}

Comparison of our thermodynamic quantities with those found semiclassically from the Euclidean path integral offers a robust check on both our results and their interpretation. In particular we will compare the thermodynamic partition function with an approximation of the Euclidean path integral partition function. Euclidean solutions of the equations of motion, also known as instantons, represent saddle points in the finite temperature gravitational path integral. Therefore their action approximates the log partition function \cite{gibbons1977action}:
\be \label{semi} -\log Z \approx I_E. \ee
Before discussing the partition function via the Euclidean path integral, we must review the geometry of the Euclidean C-metric. Performing a Wick rotation $\tau = i t$ on the Lorentzian C-metric, we obtain the Euclidean C-metric
\begin{equation} \label{EucMetric}
    ds^2 = \frac{1}{A^2(x-y)^2} \Big( - G(y) d \tau^2 - \frac{dy^2}{G(y)} + \frac{dx^2}{G(x)} + \alpha^2 G(x) d \phi^2 \Big).
\end{equation}
It is electrovacuum almost everywhere, with gauge field
\begin{equation}\label{euclideangaugepotential}
    A_\mu d x ^\mu = -iq (y + 1) d \tau.
\end{equation}
Previously $y=\zeta_2$ and $y=-1$ were horizons, but now they are more like tips of cones. In particular, with $\zeta_2 < y < -1$ the Euclidean C-metric \eqref{EucMetric} is geodesically complete. Note then $G(y) < 0$, so it is Euclidean everywhere, justifying the name. The topology of the Euclidean C-metric is $S^2 \times S^2 - \{ \mathrm{pt} \}$. One $S^2$ factor is parameterized by $(y, \tau)$ and the other by $(x, \phi)$. The removed point ``$\mathrm{pt}$'' is $x = y = -1$ and corresponds to spatial infinity. The metric retains the conical singularity sourced by the cosmic string at $x = -1$.\footnote{The conical singularity on the string must remain in the Euclidean metric because it is physically sourced by the string. This is necessary for the metric to solve the Euclidean equations of motion and thus be a saddle point of the path integral. See \cite{hawkross}.} However, since the Lorentzian C-metric has no stress-energy source on the horizons (away from the string), it seems most natural to take the stress-energy to vanish at $y=\zeta_2$ and $y=-1$ in the Euclidean solution as well (away from the string). That is, we would like to make $\tau$ periodic such that the ``cones" are actually disks. We can avoid a conical singularity at $y = \zeta_2$ if we give $\tau$ a periodicity $\tau \sim \tau + \beta$, where
\begin{equation}
    \beta = \frac{4 \pi}{|G'(\zeta_2)|} = N \frac{2 \pi }{\kap_{\rm bh}}.
\end{equation}
However, a conical singularity will remain at $y = -1$ unless we have
\begin{equation}
    |G'(\zeta_2)| = G'(-1).
\end{equation}
The above equation is called the temperature matching condition, and has been imposed in \cite{hawkross, stromzhib}. It is equivalent to
\begin{equation}
    \zeta_2 - \zeta_1 = 2
\end{equation}
assuming we require $\zeta_2 \neq -1$ so the acceleration horizon and black hole horizon remain distinct. The temperature matching condition fixes a parameter, bringing the number of free parameters of the C-metric down from three to two. It can also be rewritten as
\begin{equation}
    A = \frac{\sqrt{m^2 - q^2}}{q^2}
\end{equation}
which shows that, assuming $A$ and $m$ are finite, the temperature matching condition can only be satisfied if $q^2 > 0$, i.e. if the black holes are charged.

One final way to express the temperature matching condition is
\begin{equation}
    \kap_{\rm bh} = \kap_{\rm acc}.
\end{equation}
The physical interpretation of the temperature matching condition is that the rate at which the black hole is fed Unruh radiation exactly equals the rate at which it emits Hawking radiation, thus rendering the C-metric quantum mechanically stable.

\begin{figure}
  \centering
  \includegraphics[width=0.6\textwidth]{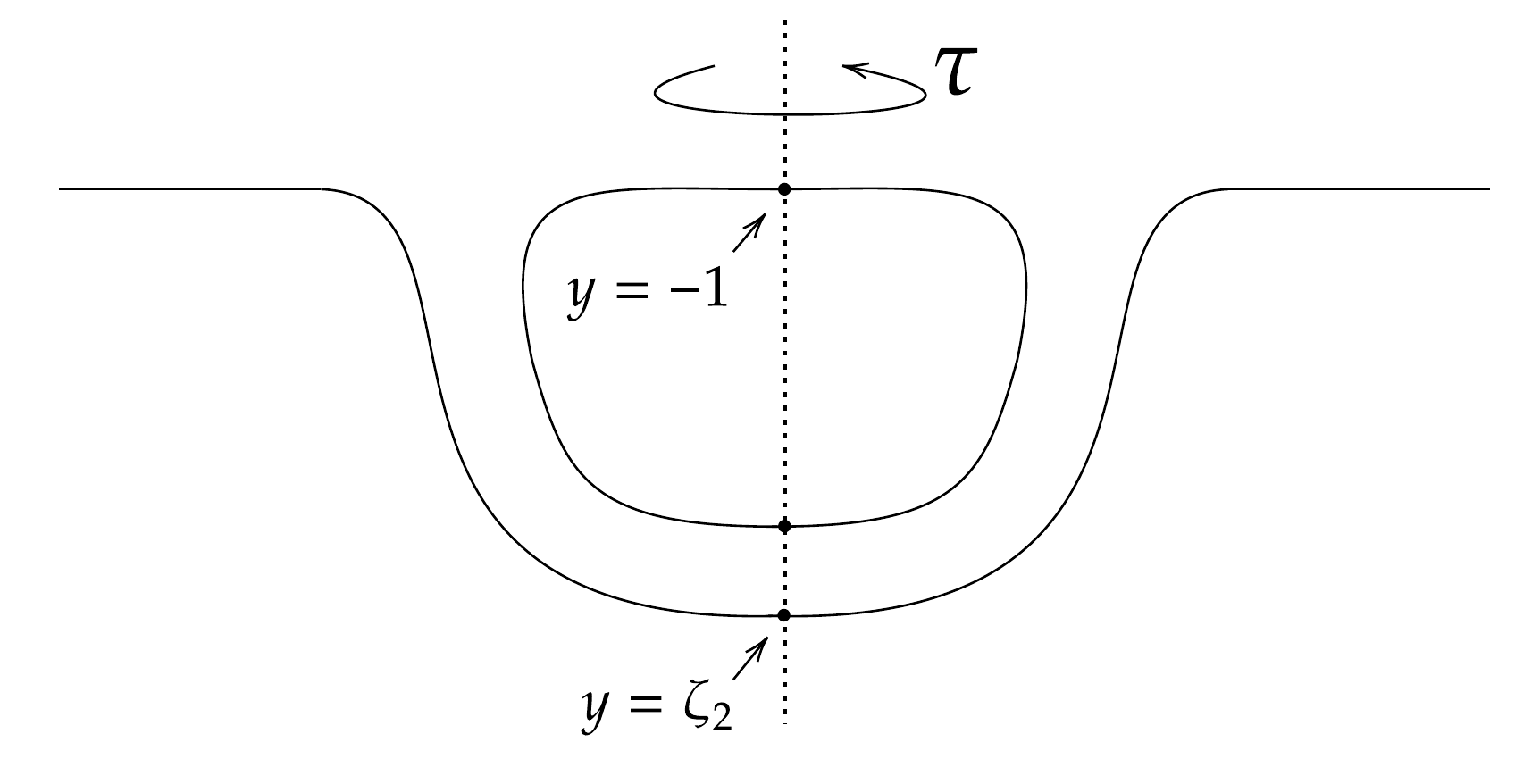}
    \caption{\label{fig:euclidean} The Euclidean C-metric can be obtained by taking a $t = 0$ slice of the Lorentzian C-metric and sweeping it around by the periodic Euclidean time $\tau$. Here we have only drawn the $x = 1, -1$ slice of the full three-dimensional $t = 0$ geometry.}
\end{figure}

The Euclidean C-metric can be pictured by taking the $t = 0$ slice of the Lorentzian C-metric and rotating it around the $y = -1$ and $y = \zeta_2$ axis by the angle $\tau$, as in figure \ref{fig:euclidean}.

We now compute the action of the Euclidean C-metric. The Euclidean action has three components, coming from the gravitational field, the electromagnetic field, and the string:
\begin{equation} \label{EucAction}
    I_E = I_E^{\rm grav} + I_E^{\rm em} + I_E^{\rm string},
\end{equation}
\begin{align}
     I_E^{\rm grav} &= -\frac{1}{16 \pi} \int d^4 x \sqrt{g} R - \frac{1}{8 \pi} \int d^3 x \sqrt{h} K, \\
     I_E^{\rm em} &= -\frac{1}{16 \pi} \int d^4 x \sqrt{g} F^{\mu \nu} F_{\mu \nu}, \\
     I_E^{\rm string} &= \mu \int d^2x \sqrt{\gamma}.
\end{align}
This action is evaluated out to a large surface approaching spatial infinity. The surface is defined by $\eps = \eps_{\rm min}$ where $\eps_{\rm min} \to 0$. $h$ is the induced metric on this surface and $K$ is its extrinsic curvature. $\gamma$ is the induced metric on the string worldsheet. As usual for a non-compact spacetime, the na\"ive action diverges so we will actually compute the difference in action relative to a background. In this case we take as our background the Euclidean cosmic string spacetime.

We begin with the gravitational term. For the C-metric, $R = 0$ everywhere except on the string where it has a delta function singularity. When integrating over a conical singularity in a two-dimensional manifold, one may use the Gauss-Bonnet theorem to show that $\int d^2 x \; \sqrt{g} R = 2 \delta_{\rm def}$, where here $\delta_{\rm def}$ is the deficit angle of the singularity. It may then be shown that the Einstein-Hilbert action of a string-like singularity in four dimensions is given by $\int d^4 x \sqrt{g} R = 2 \ca \delta_{\rm def}$, where $\ca$ is the area that the cusp sweeps out in the transverse directions. For the cosmic string in the Euclidean C-metric, this area can be expressed in terms of the thermodynamic length as
\begin{equation}
    -\frac{1}{16 \pi} \int d^4 x\sqrt{g} R = -\beta \mu \tfrac{1}{N} \ell^{(C)}.
\end{equation}
The expression for the Euclidean cosmic string background metric is analogous.

The Gibbons-Hawking-York boundary term is evaluated on the $\eps = \eps_{\rm min}$ surface where $\eps_{\rm min} \to 0$. The difference in this boundary term between the C-metric and cosmic string background is zero. Similar to the vanishing of $\slashed{\delta} \Delta M_{\rm hemi}$ in \eqref{reghemi}, this is a consequence of the agreement of the C-metric and the cosmic string background metric at subleading order in $\eps$. The total gravitational portion of the difference in Euclidean action is then
\be \Delta I_E^{\rm grav} = -\beta \mu \tfrac{1}{N} \Delta \ell. \ee

Next, the contribution from the Nambu-Goto string action may also be expressed using the thermodynamic length. One finds
\begin{equation}
    \Delta I_E^{\rm string} = \beta \mu \tfrac{1}{N} \Delta \ell.
\end{equation}
Note that this cancels out the contribution from the conical singularity sourced by the string in the Einstein-Hilbert action. A similar cancellation was noted in \cite{Almheiri:2019qdq} in the context of replica wormholes.

Finally we come to the electromagnetic term. Using the equation of motion $\nabla_\mu F^{\mu \nu} = 0$, we can write the electromagnetic Lagrangian as a total derivative: $F^{\mu \nu} F_{\mu \nu} = \nabla_\mu (2 F^{\mu \nu} A_\nu )$. Therefore the action may be expressed as a boundary integral. However, there is a small issue: the coordinate $\tau$ is ill-defined at $y = -1$ and $y = \zeta_2$. This makes the gauge potential singular at those locations unless it is zero. Our choice of gauge in \eqref{euclideangaugepotential} makes the potential vanish at $y = -1$, meaning the boundary integral at spatial infinity vanishes as well, but the gauge potential remains singular at $y = \zeta_2$. To account for this, we must create an inner boundary of integration on a surface of constant $y$ and push it up against $y = \zeta_2$. This gives a final result of
\begin{equation}
    I_E^{\rm em} = -\half \beta \tfrac{1}{N} \Phi Q = -\frac{\pi}{\kap_{\rm bh}} \Phi Q.
\end{equation}
There is no contribution to the electromagnetic portion of the action from the background cosmic string action.

The total action is then\footnote{One finds the same action by borrowing the result from \cite{hawkross} for the magnetically charged C-metric and adding the term $-\Phi Q/T$ to it, the necessity of which is explained in \cite{HawkRossDuality}.}
\begin{equation} \label{InstAction}
    \Delta I_E = -\frac{\pi}{\kap_{\rm bh}} \Phi Q.
\end{equation}

\section{The temperature-matched thermodynamic interpretation} \label{temp matched}

In our first law \eqref{introlaw} we vary three parameters. We can spend one to set $\kap_{\rm acc} = \kap_{\rm bh}$, leaving us with two.\footnote{This condition holds for black holes produced via quantum tunneling on a cosmic string \cite{hawkross}.} Now that both horizons emit radiation at the same temperature the system should be in equilibrium and we may attempt a full thermodynamic interpretation. That is, we write
\be \delta \Delta M_{\rm boost} = T \delta \Delta S + \Phi \delta Q + \Delta \ell \, \delta \mu \ee
where
\begin{align} \label{temperature} T &= \frac{\kap_{\rm acc}}{2\pi} = \frac{\kap_{\rm bh}}{2\pi}, \\
\Delta S &= \frac{1}{4} \left( \Delta \ca_{\rm acc} + \ca_{\rm bh} \right), \\
\Delta M_{\rm boost} &= 0.
\end{align}
We still have the Smarr relation
\begin{equation}
    0 = T \Delta S + \half \Phi Q.
\end{equation}
Note that $\Delta S < 0$, but as it is the change in entropy compared to the cosmic string background this is not a problem.

Using these quantities, we can write the standard thermodynamic grand potential of the system, $\Delta \Omega$, as
\be \begin{aligned}
    \Delta \Omega &= \Delta M_{\rm boost} - T\Delta S - \Phi Q \\
    & = -\half \Phi Q, \label{grandpot}
\end{aligned} \ee
where we used the Smarr relation to get the second line. This is simply related to the thermodynamic partition function by
\be -\log Z_{\rm thermo} = \Delta \Omega / T. \ee
We are now in a position to confirm that this thermodynamic partition function agrees with the semiclassical one, \eqref{semi}. The check boils down to
\begin{equation}
    \Delta I_E \stackrel{?}{=} \Delta \Omega/T.
\end{equation}
This equality does indeed hold by \eqref{InstAction}, \eqref{grandpot}, and \eqref{temperature}. This agreement with semiclassical methods offers a strong endorsement for the thermodynamic interpretation of the quantities associated with boost time.

Let us now turn to the question of what the unambiguous temperature of the C-metric is by finally choosing a normalization $N$ of the boost Killing vector $\xi$. The most natural choice matches the norm of $\xi$ to that of a boost in the cosmic string background asymptotically. This is accomplished through $N = \frac{2}{  G'(-1)}$, or equivalently $\kap_{\rm acc} = 1$.\footnote{The authors of \cite{umass} similarly required $\kap_{\rm acc} = 1$, but in their C-metric example they took $\p_t$ as their Killing vector rather than allowing for a general normalization. Accordingly they needed to fix one of their C-metric parameters to impose $\kap_{\rm acc} = 1$, rather than merely fixing the normalization as we do.} Therefore, the temperature is always
\begin{equation}
    T = \frac{1}{2 \pi}.
\end{equation}
It is dimensionless because boost time is dimensionless. It is somewhat strange to have a first law with a fixed temperature, but this can be viewed as a natural consequence of the non-compactness of the acceleration horizon. It extends far away from the black hole, where the metric is nearly flat, and so its temperature must be consistent with the flat case of Rindler. This point about fixed periodicity was also made in \cite{hawkhor}.

\section{Speculation on firewalls} \label{firewall}

In this section we speculate on how to treat temperature when the two horizons have different surface gravities. As we mentioned before, this is still an open question \cite{SdS}. Most attempts to answer it offer a single temperature \cite{Shankaranarayanan, PappasKanti, TopoTemp}. However, here we explore the idea that a continuum of temperatures may be possible. Even for a spacetime with a single horizon, it is not unheard of to have a temperature other than the Hawking temperature $T_{\rm H} = \frac{\kap}{2 \pi}$. One example is the Boulware vacuum for the Schwarzschild spacetime, which has zero temperature for static observers \cite{Candelas}. It is well-known that the stress-energy tensor in the Boulware vacuum diverges near the horizon. This signifies the existence of a firewall, which is related to the lack of entanglement across the horizon. The existence of a firewall can also be seen using the Euclidean section of the spacetime --- if we use a different periodicity for Euclidean time, there will be a conical singularity at the horizon representing a delta-function stress-energy tensor from the firewall. If we treat the singularity as a physical membrane, the action can be computed just as it was for the cosmic string.

For the C-metric, if we do not match the surface gravities of the black hole and acceleration horizons then there is necessarily a firewall on at least one horizon. We will consider an arbitrary temperature $T$, in which case there are generically firewalls on both horizons. If we introduce action terms for the firewall membranes,
\be I_E^{\rm fire, \, acc} = \sigma_{\rm acc} \int d^2x \sqrt{\gam}, \qquad I_E^{\rm fire, \, bh} = \sigma_{\rm bh} \int d^2x \sqrt{\gam}, \ee
where $\sigma_{\rm acc}$ and $\sigma_{\rm bh}$ are the firewall energy densities
\be \sigma_{\rm acc} \equiv \frac{1}{4} \left( 1 - \frac{\kap_{\rm acc}}{2\pi T} \right), \qquad \sigma_{\rm bh} \equiv \frac{1}{4} \left( 1 - \frac{\kap_{\rm bh}}{2\pi T} \right), \ee
we find that they exactly cancel the contributions of the horizon conical singularities to the Einstein-Hilbert action.
Building on the results of section \ref{Euclidean}, we therefore have
\be I_E^{\rm grav} + I_E^{\rm fire, \, acc} + I_E^{\rm fire, \, bh} + I_E^{\rm string} = 0. \ee
 The electromagnetic action is unaffected by the firewalls, and the temperature simply appears as an overall factor. The total Euclidean action is
\be I_E = I_E^{\rm em} = -\half \Phi Q / T. \ee
The grand potential is then
\be \ba \Delta \Omega = T \Delta I_E & = -\half \Phi Q \\
& = \frac{\kap_{\rm acc}}{8\pi} \Delta \ca_{\rm acc} + \frac{\kap_{\rm bh}}{8\pi} \ca_{\rm bh} \\
& = T \Delta S - T \sigma_{\rm acc} \Delta \ca_{\rm acc} - T \sigma_{\rm bh} \ca_{\rm bh}. \ea \ee
We assume the entropy is still $\Delta S = \frac{1}{4} (\Delta \ca_{\rm acc} + \ca_{\rm bh})$. Using these definitions, the first law \eqref{thefirst} can be rewritten as
\be 0 = T \delta \Delta S - T \sigma_{\rm acc} \delta \Delta \ca_{\rm acc} - T \sigma_{\rm bh} \delta \ca_{\rm bh} + \Phi \delta Q + \Delta \ell \, \delta \mu. \ee
In this form there is a single temperature $T$ along with energy contributions from the new firewalls.

\section{Conclusion}

We studied the thermodynamics of the C-metric with respect to canonical charges defined by boost time. As the C-metric has two horizons and a cosmic string, we drew a distinction between global and local first laws. We derived a global first law for the spacetime in terms of a change in boost mass $\Delta M_{\rm boost}$ which we found to vanish. We then showed that the local first law of the black hole horizon reproduced the Reissner-Nordstrom first law in the small string tension limit. Upon matching the temperatures of the acceleration horizon and the black hole horizon, we were able to assign traditional thermodynamic interpretations to the quantities in the global first law. We showed agreement between the thermodynamic and path integral partition functions which served as an independent check on the thermodynamic interpretation. For all of these reasons, we believe that \eqref{introlaw} can rightfully be called the first law of the C-metric.

\acknowledgments

We are grateful to Andy Strominger for acquainting us with the C-metric, and for his generous guidance. We also thank Kevin Nguyen, Jakob Salzer, and Rudranil Basu for useful discussions. AB gratefully acknowledges support from NSF grant 1707938 and the Fundamental Laws Initiative. NM gratefully acknowledges support from NSF GRFP grant DGE1745303.

\bibliography{cthermo}
\bibliographystyle{utphys}

\end{document}